\documentstyle[preprint,aps]{revtex}

\def\addcontentsline#1#2#3{\relax}
\renewcommand{\baselinestretch}{1}
\begin{document}
\draft
\title{ Adiabatic connection between the
  RVB  State
and the ground state of the  half filled periodic Anderson model}
\author{K. Kimura$^{1}$$^*$, Y. Hatsugai $^{2}$, and M. Kohmoto$^{1}$}
\address{
${}^1$Institute for Solid State Physics, University of Tokyo,
7-22-1, Roppongi, Minato-ku, Tokyo 106, Japan
}
\address{
${}^2$Department of Applied Physics, University of Tokyo, 7-3-1, Hongo,
Bunkyo-ku, Tokyo 113, Japan
}
\maketitle
\begin{abstract}
   A one-parameter family of models that interpolates between the periodic A
nderson
model with infinite  repulsion at  half-filling and a model whose
ground state is exactly the  Resonating-Valence-Bond  state is studied.
It is shown  numerically that
the  excitation gap does not collapse. Therefore the ground states of the
two models
are adiabatically connected.
\end{abstract}

\newpage
\section{Introduction}
\label{sec:1}
Recently correlation effects in  electronic systems have been focused
and studied extensively.
This is an old problem, however, it still is supplying interesting new
physics both experimentally and theoretically.

One can classify the ground states of strongly correlated systems into two.
The one
is a metallic state which  have a gapless excitation.
The Fermi-liquids  and the Tomonaga-Luttinger liquids in  one-dimension
are in this class.
The other is an insulator which  has a finite excitation gap. A  simple example
is a band insulator. Also there is another type of  insulators which are
caused by
correlation (Mott-insulators).
A well-known example of the correlated insulators is the half filled Hubbard
model in
one dimension.

 Another  example with the correlation gap is an energy gap of the half-filled
Kondo lattice in one dimension. The charge degree of freedom  on the sites with
on-site  Coulomb repulsion are frozen.  In this model both  the charge
and the spin degree of freedom have a finite excitation gap though the
lowest one is the
spin excitation\cite{ueda}. The periodic Anderson model which we
investigate  is a model
where the charge degree of freedom are also  active.

Principle of  adiabatic continuation is  important
 in   condensed matter physics. For example,
 the basic assumption of the Fermi liquid theory is that the interacting system
with quasiparticles is adiabatically connected to the non-interacting system
with
several phenomenological parameters.  More specifically, the
non-interacting fermions has one to one correspondence to the quasiparticle
of the
interacting electrons and there are no gap closing in the process of
 increasing the interaction from zero to reach the interacting model.
Another notable
example is  the theory of the fractional quantum Hall effect.  The  adiabatic
transformation in which the external magnetic fluxes are  put on the
electrons to
become bosons \cite {Zhang} or composite fermions\cite{jain} is  the crucial
assumption.

In this paper we choose the model of Strack\cite{str} in one dimension as
the canonical
system with the correlation gap.
The ground state is exactly the Resonating-Valence-Bond (RVB)
state\cite{str,Tas1,Tas2,bares,yamanaka}. See, {\it e.g.}, Ref.9 for the
form of the RVB
state. In this model, some of the correlation functions are obtained
exactly\cite{bares,yamanaka}. Moreover it is connected to the periodic
Anderson model in
one dimension as a parameter is varied. The periodic Anderson model has been
commonly used
to describe the correlation effects in   heavy fermion compounds and it
reduces to the
Kondo lattice model  when the valence fluctuation is prohibited\cite{ueda}.
In order to
clarify the relation between the two model  we numerically obtain the ground
state energy
and the excitation gap for  intermediate Hamiltonians.

\section {Models}

The Hamiltonian of the Strack model is
\begin{eqnarray}
H_{ST} &=& \wp\sum_{n,\sigma} \{  (-\lambda_{1}
\lambda_{2}c_{n+1,\sigma}^{\dag}c_{n,\sigma}-\lambda_{1}c_{n+1,\sigma}
^{\dag}f_{n,\sigma}-\lambda_{2}
c_{n,\sigma}^{\dag}f_{n,\sigma}+h.c. \nonumber\\
 &+& \epsilon^{c}c_{n,\sigma}^{\dag}c_{n,\sigma}+
\epsilon^{f}f_{n,\sigma}^{\dag}f_{n,\sigma}
\} \wp,
\label{hst}
\end{eqnarray}
where   ${n}$ is an index of the unit cell. In  Fig.1(a), the lattice
structure of
the model is shown where
  ${\circ}$ and  ${\bullet}$ denote $f$- and  ${c}$- sites
respectively.
Electrons at ${f}$-sites feels infinitely
large on-site Coulomb repulsion(${U=\infty}$)  and  ${c}$-sites do
not have Coulomb repulsion(${U=0}$).
The projection operator $\wp$ represents to project out
the states with doubly occupancy at the  $f$-sites.
 When one imposes  ${\epsilon^{c}=2-(\lambda_{1}^{2}+
\lambda_{2}^{2})\ }$, and ${\epsilon^{f}=2-2=0}$,
the ground state  at  half-filling  is explicitly written
\begin{eqnarray}
|\Phi_{G}\rangle & = &
\wp\prod_{n,\sigma}
(\lambda_{1}c_{n,\sigma}^{\dag}+\lambda_{2}c_{n+1,\sigma}^{\dag}
+f_{n,\sigma}^{\dag})|0\rangle. \nonumber \\
 & = &
\prod_{n}
(\lambda_{1}\lambda_{2}d_{c_{n},c_{n+1}}^{\dag}+
\lambda_{1}^{2}d_{c_{n},c_{n}}^{\dag}  \nonumber \\
&+&
\lambda_{2}^{2}d_{c_{n+1},c_{n+1}}^{\dag}+
\lambda_{1}d_{c_{n},f_{n}}^{\dag}+
\lambda_{2}d_{f_{n},f_{n+1}}^{\dag})|0\rangle,
\label{gd1}
\end{eqnarray}
where
\begin{equation}
d_{\alpha_{i},\beta_{j}}^{\dag}= \left\{
 \begin{array}{ll}
\alpha_{i,\uparrow}^{\dag}\beta_{j,\downarrow}^{\dag}+
\alpha_{j,\uparrow}^{\dag}\beta_{i,\downarrow}^{\dag}
  \quad \mbox{ for\  $ i\neq j$ } \\
\alpha_{i,\uparrow}^{\dag}\alpha_{i,\downarrow}^{\dag}
         \quad \mbox{ for $i = j$ } .
  \end{array}\right.
\label{pair}
\end{equation}
Thus it is given by  creations of  nearest-neighbor singlet pairs on the
vacuum.
 This state is the RVB  state
which we use as the canonical ground state with the correlation gap.

 The existence of the
finite energy gap has not been shown analytically. But it is numerically
confirmed in the
present work. This is consistent with the behavior of
correlation functions of a local quantities which are analytically shown to be
exponentially decaying \cite{bares,yamanaka}.
 One can expect that the excitation above the ground state is closely
related to a local singlet-triplet excitation which apparently has a finite
energy cost.

The Hamiltonian of the periodic Anderson model is
written
\begin{eqnarray}
H_{PA} =  t\sum_{n,\sigma}(c_{n+1,\sigma}^{\dag}c_{n,\sigma}+h.c.)+
V\sum_{n,\sigma}(c_{n,\sigma}^{\dag}f_{n,\sigma} + h.c.)+ \nonumber\\
\epsilon^{f}\sum_{n,\sigma}f_{n,\sigma}^{\dag}f_{n,\sigma} + U
\sum_{n}f_{n,\uparrow}^{\dag}f_{n,\uparrow}
f_{n,\downarrow}^{\dag}f_{n,\downarrow},
\label{pa}
\end{eqnarray}
 where ${U}$ is the on-site Coulomb repulsion on ${f}$-sites.
 We consider the strong coupling limit  $U\to \infty$.

These two Hamiltonians (\ref{hst}) and (\ref {pa}) is connected by changing
hopping
elements of the Strack model as shown in the Fig.1(c).

The  intermediate Hamiltonian  we study is
\begin{eqnarray}
H_{C} = \wp\sum_{n,\sigma }
[( t c_{n+1,\sigma}^{\dag}c_{n,\sigma} + t_{1}c_{n+1,\sigma}
^{\dag}f_{n,\sigma} + t_{2}
c_{n,\sigma}^{\dag}f_{n,\sigma}+h.c.)\nonumber\\
 + \epsilon^{c}c_{n,\sigma}^{\dag}c_{n,\sigma}+
\epsilon^{f}f_{n,\sigma}^{\dag}f_{n,\sigma}]\wp.
\label{rel}
\end{eqnarray}
The Strack Hamitonian (\ref{hst}) which is given by setting
$t_1$ and $t_2$    as  $ t = -\lambda_{1}\lambda_{2}$,  $t_{1}
= -\lambda_{2},  t_{2} =
  -\lambda_{1} $.
Also when ${t_{1} = 0}$, it reduces to the periodic
Anderson model (\ref{pa}) with ${ U = \infty}$.

\section{Numerical Results}
To calculate the ground states and the
energy gaps for  sufficiently large systems, we used the White's method (DMRG)
\cite{Whi1,Whi2}. Also numerical diagonalizations was performed for
relatively small
systems  to check the validity of the DMRG results.
It is interesting to note that DMRG is exact in the Strack
model\cite{kimura}. This fact
supports the locality of the RVB state.

First we start with the Strack model by setting $t=t_{1} =t_2= -1 $ and
$\epsilon^c=\epsilon^f=0
$ in (\ref{rel}). Then it is identical to the Strack's Hamiltonian
(\ref{hst})  with
${\lambda_{1}=\lambda_{2}= 1}$.
By changing  $t_1$ while keeping the other parameters fixed in $H_C$,
one gets the
periodic Anderson model with $\epsilon^f=0$ when $t_1=0$.

The excitation gap obtained numerically are plotted in Fig. 2. It interpolates
${t_1=-1}$
(the Strack model) and ${t_1=0}$ (the periodic Anderson model).
We used a periodic boundary condition
and each values are calculated by taking an extrapolation to the infinite
system
size.
 As a reference,  the energy gap obtained by the slave boson
method is also plotted \cite{kimura}.

The system size dependence of the energy gap is shown in Fig.3 with the
results
with open boundary condition of the periodic Anderson model ($t_1=0$ in
(\ref{rel}) )

As shown in  Fig.2, the excitation gap of the
half-filled periodic Anderson model is connected
to that of the Strack's model without  gap closing.
It implies that the ground state of the periodic Anderson model at half
filling
may have close connection to that of the RVB state.
For example, both ground states are singlet and have local nature. The
excitations are
expected to be closely related to local singlet-triplet excitations

\renewcommand{\baselinestretch}{1.0}

\begin{figure}
Fig. 1
Lattice structure :(a) The Strack model, (b) the periodic Anderson model,
and (c) an intermediate model  connecting  (a) and (b).

Fig. 2
Excitation gap versus $t_{1}$.

Fig. 3
Excitation gap versus $1/L$ : $L$ is the system size.

\end{figure}


\begin{references}
\bibitem[*]{} Present address: The Central Research Laboratory, Canon Co. ,
Atsugi-city, Kanagawa-prefecture, Japan.

\bibitem{ueda}K. Tsunetsugu, Y. Hatsugai, K. Ueda, and M. Sigirist,
Phys. Rev. B{\bf 46} 3175 (1992),
Tsunetsugu and K. Ueda,
Rev. Mod. Phys.to be published (1997).

\bibitem{Zhang} S.C. Zhang, T.H. Hansson, and S. Kivelson,
 Phys. Rev. Lett. {\bf 62}, 82 (1989).

\bibitem{jain} J.K. Jain, Phys. Rev.~ Lett. {\bf 63}, 199 (1989).

\bibitem{str}R. Strack, Phys. Rev. Lett. {\bf70}, 833 (1993).

\bibitem{Tas1}H. Tasaki, Phys. Rev. Lett. {\bf 70},3303 (1993).

\bibitem{Tas2}H. Tasaki, Phys. Rev. B. {\bf 49}, 7763 (1993).

\bibitem{bares} P. A. Bares and P. A. Lee, Phys. Rev. B {\bf 49}, 8882
  (1993).

\bibitem{yamanaka}M. Yamanaka, S. Honjo, Y. Hatugai, and M. Kohmoto,
 J. Stat. Phys. {\bf 84}, 1133 (1996).

\bibitem{kohmoto} M. Kohmoto,  Phys. Rev. B37, 3812 (1988).


\bibitem{Whi1}S. R. White, Phys. Rev. Lett. {\bf 68}, 3487 (1992).

\bibitem{Whi2}S. R. White, Phys. Rev. B. {\bf 48}, 10345 (1993).

\bibitem{BG}U. Brandt and A. Giesekus, Phys. Rev. Lett. {\bf68}, 2648
  (1992)

\bibitem{kimura} Master Theses, University of Tokyo (1996) unpublished.



\end{references}
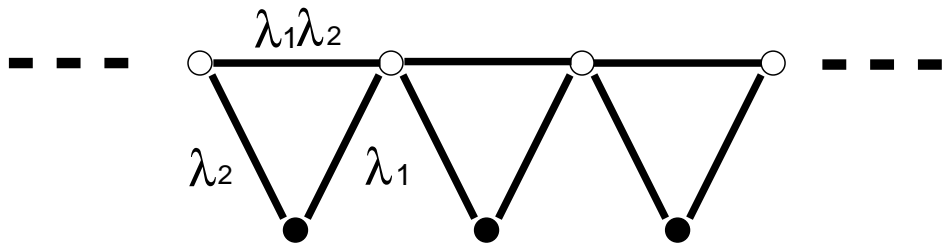
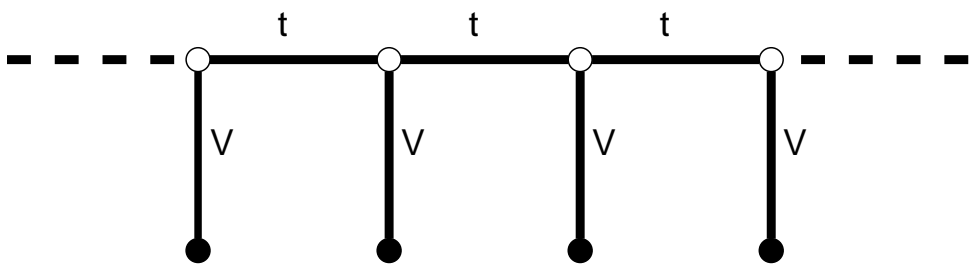
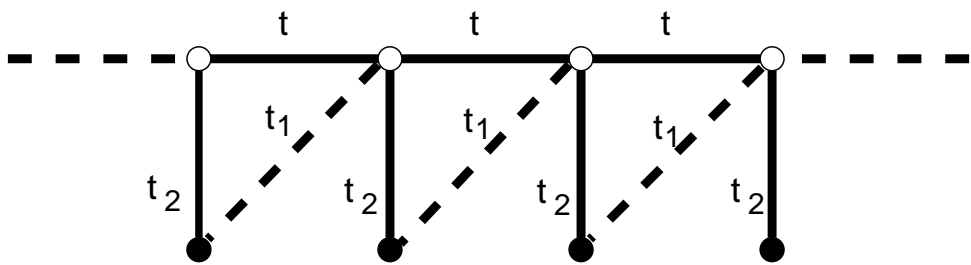
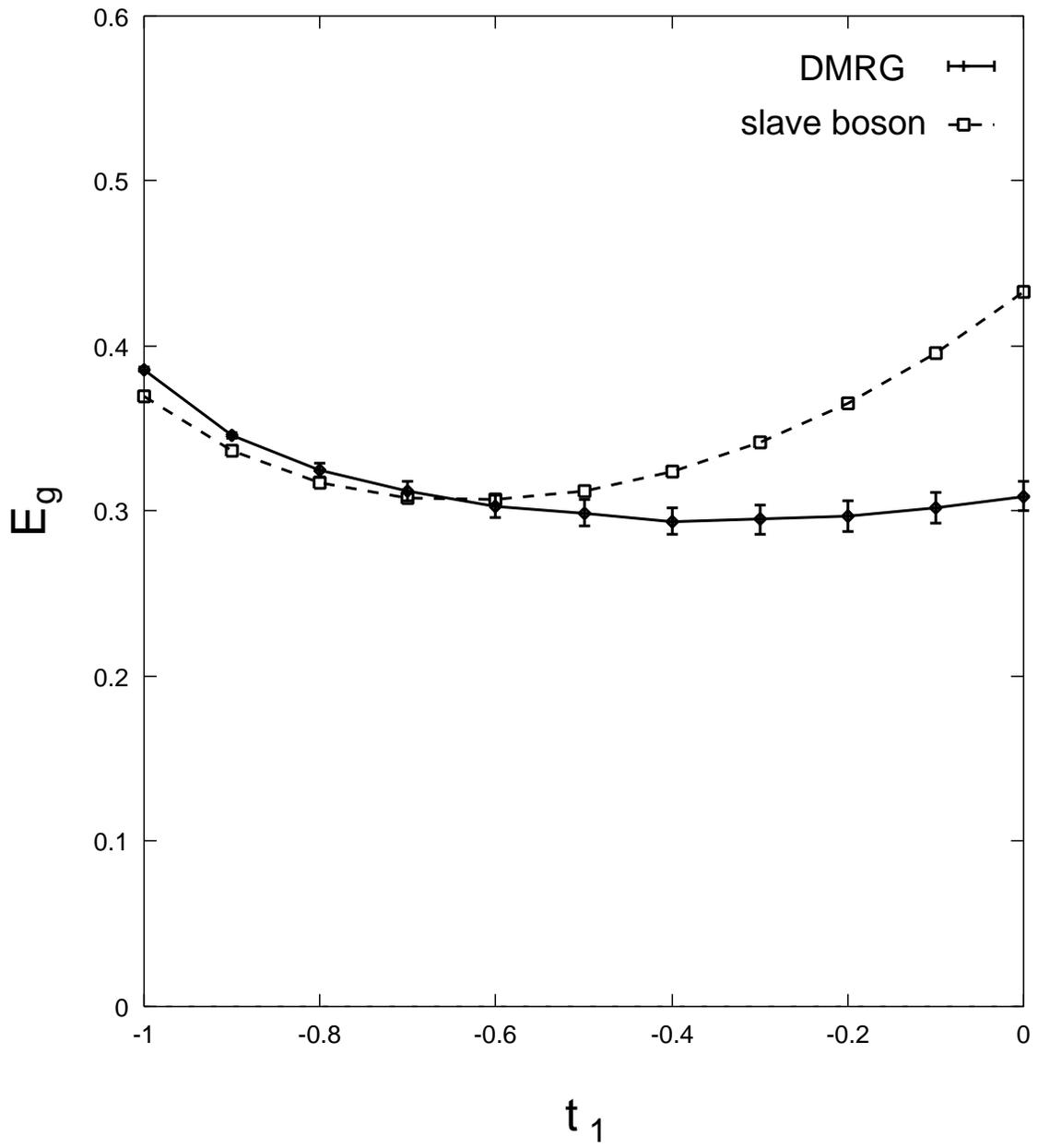
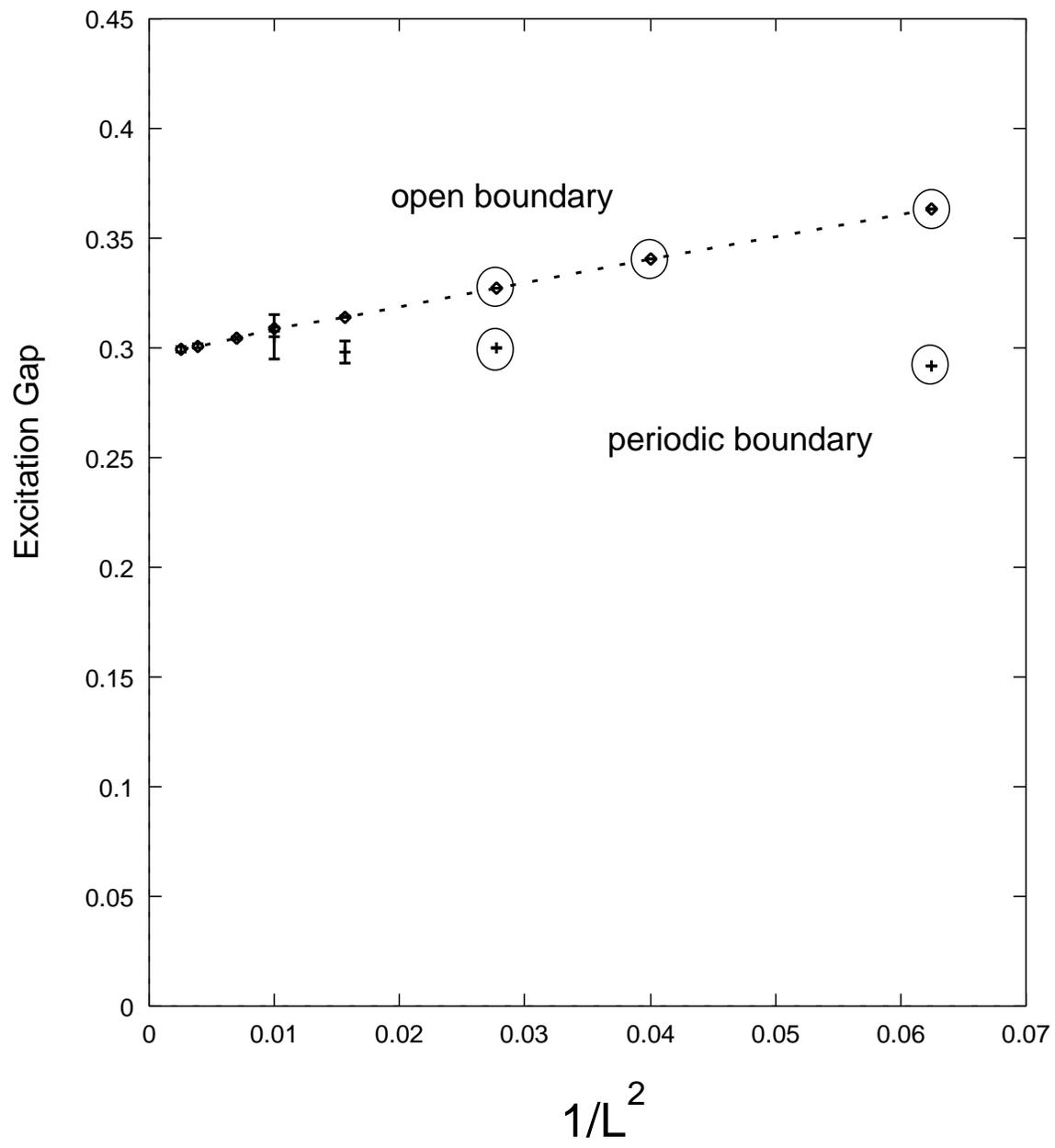
\end{document}